\documentclass{article}

\usepackage{amsmath}
\usepackage{mathtools}
\usepackage{amsfonts}
\usepackage{amssymb}
\usepackage{fancyhdr}
\usepackage[margin=1in]{geometry}
\usepackage{graphicx}
\usepackage{nicefrac}
\usepackage{bm}
\usepackage{xcolor}
\usepackage[format=plain,indention=.5cm, font={it,small},labelfont=bf]{caption}
\usepackage{subcaption}
\usepackage{float}
\usepackage{enumerate}
\usepackage{tikz}
\usetikzlibrary{arrows.meta}
\usepackage[all]{xy}
\usepackage{wrapfig}
\usepackage[nodisplayskipstretch]{setspace}
\usepackage{multicol}
\usepackage{appendix}

\usepackage[colorlinks = true,linkcolor = blue,citecolor = blue]{hyperref}

\usepackage{cleveref}
\crefname{subsection}{}{subsections}

\newcommand{\nhalf}{\nicefrac{1}{2}}

\renewcommand{\b}{\bm}

\definecolor{dgreen}{RGB}{49,128,23}
\definecolor{nicepink}{RGB}{255, 0, 102}
\definecolor{nicered}{RGB}{255, 80, 80}

\newcommand{\bR}{\mathbb{R}}

\newcommand{\cQ}{\mathcal{Q}}

\newcommand{\ka}{\kappa}

\begin{document}
\title{Metabolite mediated modeling of microbial community dynamics captures emergent behavior more effectively than species-species modeling.}

\author{
J.D. Brunner$^{1,2}$ and N. Chia$^{1,2}$}
\date{}





\maketitle

 {\scriptsize \noindent$^{1}$Division of Surgical Research, Department of Surgery, Mayo Clinic\\
 $^{2}$Microbiome Program, Center for Individualized Medicine, Mayo Clinic }
\begin{abstract}
Personalized models of the gut microbiome are valuable for disease prevention and treatment. For this, one requires a mathematical model that predicts microbial community composition and the emergent behavior of microbial communities. We seek a modeling strategy that can capture emergent behavior when built from sets of universal individual interactions. Our investigation reveals that species-metabolite interaction modeling is better able to capture emergent behavior in community composition dynamics than direct species-species modeling.

Using publicly available data, we examine the ability of species-species models and species-metabolite models to predict trio growth experiments from the outcomes of pair growth experiments. We compare quadratic species-species interaction models and quadratic species-metabolite interaction models, and conclude that only species-metabolite models have the necessary complexity to to explain a wide variety of interdependent growth outcomes. We also show that general species-species interaction models cannot match patterns observed in community growth dynamics, whereas species-metabolite models can. We conclude that species-metabolite modeling will be important in the development of accurate, clinically useful models of microbial communities. 
\end{abstract}

\maketitle

%

\section{Introduction}

The microbial communities of the human body, collectively called the ``human microbiome", act on the host in a symbiotic relationship which can have profound effects on health and disease \cite{braundmeier2015,calcinotto2018,walsh2019,Hale2018,flemer2017,ng2013,round2009}. This can be seen in the impact of bacterial colonization on the development of the adaptive immune system \cite{round2009} as well as the many observed microbiome alternations in diseases ranging from multiple sclerosis \cite{chen2016} to colorectal cancer\cite{Hale2018}. Importantly, these changes go beyond the presence or absence of a single species, but derive from more complex shifts in community composition. One prime example of this is the presence of pathogenic bacteria among the microbiota of disease-free asymptomatic individuals \cite{round2009}. It is clearly not enough to identify and target a single species when attempting to explain the impact of the microbiome on host health. Instead, we must understand the interactions within a microbial community. These interactions determine whether a potentially pathogenic bacteria behaves as a beneficial, neutral, or pathogenic member of the microbial community. Because these properties arise from the broader community, rather than the individual species, we call these emergent properties. Identifying and predicting microbial community composition and the resultant emergent properties are an important part of disease prevention, diagnosis, and treatment in the burgeoning world of data-driven and individualized medicine.

In order to understand the composition of the microbiome, we must understand the dynamic process of growth, invasion, and extinction which leads to a stable microbiome in healthy individuals, as well as the dynamic responses of the microbiome to changes in host health and diet. Such an understanding would give us the fundamental rules by which to alter the microbiome in an effective and stable way. We may then prevent disease and improve disease outcome by formulating these rules into a model of how microbiome composition changes in response to perturbation, and use this model to design treatments to manipulate the microbiome. 

The goal of this manuscript is to identify a modeling framework for recapitulating community growth dynamics from sets of fundamental interactions. This modeling framework must be extensible, that is allow us to directly combine models of smaller communities to create models of composite communities without discovering new parameters. As part of this goal, we would like this model to be ``as simple as possible but no simpler". These properties would allow us to generate a clinically useful dynamical model of the microbiome---one that can predict community composition dynamics from individualized information such as initial composition and environmental perturbations due to treatment. It is worth noting that in this setting, a single patient cannot in general provide enough data to accurately parameterize a purpose-built model. Instead, an individualized model needs to be built using parameters generated from previous experiments or other data-sets. 

This contradiction can be overcome by using a modeling framework that infers whole community dynamics from simple, fundamental interactions that are assumed to be universal. Universal building block interactions could be used to build purpose-built predictive dynamic models in an ``n of one" manner, meaning built with data from a single patient \cite{guyatt1990n,lillie2011n}.  It is as yet an open question as to the nature of these building blocks, and indeed if any can be found \cite{momeni2017lotka,wang2019evidence,erez2019nutrient,goyal2018,goyal2017microbial}. Here, we examine two popular modeling frameworks, species-species interaction (SSI) and species-metabolite interaction (SMI) models, and evaluate them using interdependent growth experiments of single species, pairs, and trios from Friedman et al.\cite{gore2017}. These growth experiments were carried out on flat bottom plates with an experimental growth media and serial dilution. We assume well mixed, spatially homogeneous interactions and growth. We are therefore testing this on the simplest possible situation - pair and trio growth experiments, asking whether or not a modeling strategy can recapitulate the observed outcomes of these experiments.

Perhaps the most popular candidate for a set of fundamental modeling building blocks is the set of interactions between species of microbes\cite{gore2017,mounier2008,fisher2014,stein2013ecological,kuntal2019web,angulo2019theoretical}. We call such models species-species interaction (SSI) models. This strategy follows from microbial co-occurrence networks, which can be inferred from 16s rRNA gene or metagenomic sequencing data \cite{faust2012,muller2008}.  Focusing on species-species interactions is notably the strategy of the popular Lotka-Volterra (LV) model and its generalizations \cite{gore2017,mounier2008,fisher2014,stein2013ecological}. The Lotka-Volterra model reproduces the dynamics of interacting species according the law of mass action \cite{feinberg1979,yu2018}. Therefore, it is an appropriate model of direct interaction between species in a well mixed and stable environment. The Lotka-Volterra model and SSI models in general may therefore be useful when fit to stable environments. 

SSI models can capture some of the emergent behavior of composition dynamics, but fail to capture higher order interactions which require more than two species \cite{Billick1994,ekesh}. We find in general that SSI models imply a strict condition on growth dynamics that is not observed in data. Furthermore, we find that the quadratic SSI model, usually called the ``generalized Lotka-Volterra (gLV) model", is not capable of recapitulating the entire set of pair and trio growth outcomes using a single parameterization. Although the gLV model may be useful when fit to whole communities \cite{stein2013ecological,kuntal2019web,angulo2019theoretical}, our work suggests that this model lacks the necessary complexity to be predictive when built from building blocks assumed to be universal.


Alternatively, the microbiome can be modeled by species-metabolite interaction (SMI) models, which are constructed using the interactions of individual microbes with a shared metabolite pool \cite{Niehaus2018,Posfai2017,sung2017,goyal2018,momeni2017lotka}. SMI models follow from networks which include both microbiota and metabolites, which may be inferred from literature \cite{sung2017} or from genome-scale metabolic networks\cite{chan2017}. Recently, Momeni et al.\cite{momeni2017lotka} proved that SMI models are strictly more complex than SSI models. Like SSI models, SMI models may include arbitrary complexity in interaction terms; saturating kinetics (e.g. Michaelis-Menten or Hill kinetics) are a particularly common choice \cite{hart2018microscopy,Niehaus2018,Posfai2017}. We show that a simpler quadratic species-metabolite interaction (QSMI) model can recapitulate the growth experiment outcomes of Friedman et al.\cite{gore2017} with a single parameterization.

It is worth noting explicitly that our work does not explore the accuracy of specific modeling tools \cite{Diener2018,henry2010,CobraPy,rottjers2018}, but instead examines whether the mathematical formulation of the model could ever be used to recapitulate the biological dynamics. A positive answer means that the mathematical form of the model can potentially be useful, indicating promise for future endeavors, while a negative answer indicates that the basic formulation of the model is inappropriate for predictive models of microbial communities. More specifically, we investigate this question by inspecting to what extent models of simple communities can be used to build accurate models of larger communities. Precisely, we asked whether these models have the capacity for a parameterization that recapitulates qualitative outcome of both pair and trio growth experiments. 

%

\section{Model definitions and background}
\subsection{Species-Species Interaction (SSI) models}
Species-Species Interaction (SSI) models are dynamical models of the composition of a community of organisms $i,j,k$, etc., built by assuming direct interaction between species. We model this by assuming that the population size of some species changes as the product of some per-organism growth rate and the current population size. This per-organism growth rate is then determined by interactions with other species. 

SSI models are popular in ecology, including in the study of the microbiome, because they are computationally simple to create and analyze \cite{gore2017,mounier2008,fisher2014,kuntal2019web,mougi2012diversity,thebault2010stability,allesina2012stability}. They are often fitted to large communities of microbiota, and interactions between species are assumed from this fitting \cite{stein2013ecological,angulo2019theoretical}. This fitting implies a relationship between species, and these relationships are often classified as competitive, mutual, or parasitic \cite{dohlman2019mapping}. In the study of the human microbiome, discovering interactions between species is an active area of research \cite{dohlman2019mapping}, and automated tools \cite{fisher2014,shaw2016metamis,kuntal2019web} for the construction of SSI models are used \cite{fisher2014,chen2017microbiome,shaw2017inferring,dvzunkova2018oxidative}. It is therefore of interest to understand whether or not these interactions are preserved as the community changes, and so can be used as universal building blocks.

The general form of an SSI model is as follows:
\begin{equation}\label{genLV}
\frac{dx_i}{dt} = x_i\left(f_i(x_i) + \sum_{j\neq i} h_{ij}(x_i,x_j)\right)
\end{equation}
where $x_i$ represents the biomass of organism $i$ and the functions $f_i(x_i)$ and $h_{ij}(x_i,x_j)$ respectively represent lone growth of organism $i$ and the effect of organism $j$ on growth of organism $i$. Note that we allow complete generality, including the case of distinct functions $h_{ij}$ for each pair $i,j$. It is convenient to write SSI models as product of current population size $x_i$ and a per organism growth rate $\left(f_i(x_i) + \sum_{j\neq i} h_{ij}(x_i,x_j)\right)$. The number of interaction terms in an SSI model scales with the number of pairs of $N$ microbes, and so scales as $N^2-N$.

In this manuscript, we require that $h_{ij}(x_i,x_j)$ satisfy $h_{ij}(x_i,0) =  0$ and that $h_{ij}(x_i,x_j)$ not switch sign (i.e. $h_{ij}(x_i,x_j)\geq 0$ or  $h_{ij}(x_i,x_j) \leq 0$ for any non-negative population sizes $x_i,x_j$). This simply means that species $j$ must be present to have an effect on the growth of $i$, and that species $j$ either increases the growth of species $i$ or decreases this growth (or has no effect), regardless of the population sizes of species $i$ and species $j$, while the strength of this effect may depend on population sizes.

\subsubsection*{The generalized Lotka-Volterra (gLV) model}
The simplest SSI model assumes direct, pairwise interactions proportional to species concentration, and is called the \emph{generalized Lotka-Volterra (gLV) model}.  In this model, the per-organism growth rate of a population changes proportionally to the size of each other population. Such changes may be positive or negative, indicating mutualism, parasitism, predation and competition.

The generalized Lotka-Volterra model faithfully models direct pairwise interactions between agents (e.g. physically interacting organisms or reactants in an industrial reactor) under the assumption of mass action kinetics \cite{mfeinlec,yu2018}, but does not include any possible environmental variation in interaction. It is therefore a fair representation of species-species interaction in a controlled environment. Because of this, the generalized Lotka-Volterra model is well studied and its parameters are often inferred from correlations seen in available 16S rRNA gene or metagenomic sequencing data \cite{faust2012,kuntal2019web}. Furthermore, the generalized Lotka-Volterra model is commonly used to infer relationships between species and model the microbiome \cite{mougi2012diversity,thebault2010stability,allesina2012stability,fisher2014,chen2017microbiome,shaw2017inferring,dvzunkova2018oxidative}.

This model is typically written \cite{ekesh,gore2017} as follows:
\begin{equation}\label{LV}
\frac{dx_i}{dt} = r_i x_i \left(1 - \frac{1}{K_i}x_i + \sum_{j\neq i} \alpha_{ij} x_j\right)
\end{equation}
where $r_i>0$ is the intrinsic growth rate of the community of organism $i$, $K_i > 0$ is the carrying capacity of the environment for the organism, and $\alpha_{ij}$ is the interaction between species.  Although pairwise and linear in per organism growth rate, this model can display a wide range of behaviors seen in nature, including invasion, competitive exclusion, coexistence, and multi-stability (i.e., stable long-term outcomes that are dependent on initial community sizes even for fixed parameters) \cite{gause1934,ekesh}.

\subsection{Species-Metabolite Interaction (SMI) models}
Species-metabolite Interaction (SMI) models, also called metabolite (or resource) mediated models, use the interaction of a microbe with an environmentally available metabolite as their fundamental interaction\cite{Niehaus2018,goyal2018}. Similar to SSI models, we assume that the population size of some species changes as the product of some per-organism growth rate and the current population size. In SMI models, per-organism growth rate depends on available metabolites rather than a fixed carrying capacity. Additionally, metabolites are used and produced by individual species, in many cases as a by-product of some process involving another metabolite. Interactions between microbes are then possible through manipulation of the shared metabolite pool. 

SMI models have recently become of interest in the study of the human microbiome as data on the metabolite pool has become available \cite{hart2018microscopy,watrous2012mass,perez2013gut}. Techniques for integrating species abundance and metabolomic data are now being developed to understand the mechanisms of microbiota organization \cite{noecker2016metabolic,dohlman2019mapping}. 


We define a SMI model to be any model of the type 
\begin{align}
    \frac{dx_i}{dt} &= x_i \sum_{j=1}^m f_{j}^i(x_i,y_j)\\
    \frac{dy_j}{dt} &=  \sum_{i = 1}^n \sum_{l=1}^m h_{il}^j(x_i,y_l)
\end{align}
where $n$ is the number of microbial species and $m$ the number of metabolites modeled, $x_i$ again represents biomass of organism $i$, and $y_j$ is the concentration or biomass of metabolite $j$. As species and metabolites are added to the model, the number of terms could grow as large as $NM + NM^2$, where $N$ is the number of species and $M$ the number of metabolites included.

The simplest SMI model has competition for resources, leading to the well known competitive exclusion principle which states that an environment cannot support more species than it has food sources. However, metabolite mediated models have more versatility, and do not need to hold to that principle \cite{Posfai2017}. 

Popular choices of interaction terms $f_{ij}(x_i,y_j)$ and $h_{il}^j(x_i,y_l)$ include sigmoidal (i.e. saturating) kinetics, such as Michaelis-Menten or Hill kinetics \cite{Posfai2017,hart2018microscopy}, which have a diminishing change in effect as concentrations of metabolite increase, and polynomial kinetics.

\subsubsection*{The Quadratic Species-Metabolite Interaction (QSMI) model}
The simplest SMI model is quadratic, and includes consumption and production of metabolites by microbes. This model assumes that per-organism growth rate of microbiota changes proportionally to the amount of each metabolite that it interacts with.

Similar to the gLV model, the QSMI model is a faithful model of interacting components of a well mixed system \cite{mfeinlec,yu2018}, without allowing for any variation in environmental variables. In order to model stable growing communities, we include constant dilution of microbiota as well as metabolites. The QSMI model is the closest analogue among SMI models to the generalized Lotka-Volterra model, as both are quadratic polynomials that faithfully model well-mixed interacting actors.

This gives the general model for $n$ species and $m$ metabolites
\begin{align}
    \frac{dx_i}{dt} &= x_i\left(\sum_{j=1}^m\psi_{ij}y_j - d_i\right) \label{poscfdx}\\
    \frac{dy_j}{dt} &= f_{j}  - d^*_j y_j - y_j\sum_{i =1}^n \ka_{ij}x_i + \sum_{i=1}^n\sum_{l=1}^m \phi_{il}^{j}x_iy_l \label{poscfdy}
\end{align}
where a standard competition model would have $\ka_{ij} = \psi_{ij} \geq 0$. This model assumes that a microbial population's growth rate depends on the availability of the resources being metabolized for growth, and these resources are depleted as this growth happens. The terms $\phi_{il}^jx_i y_l$ allow for the possibility that metabolites are produced as by-product of some microbial metabolic pathway.

%

\section{Results}

Recall that our goal is a model of a composite community that can be built by directly combining models of smaller communities without discovering new parameters. We therefore analyze SSI, gLV, SMI, and QSMI models with the goal of demonstrating which of these frameworks has the capacity for such a model. We find that gLV, and even general SSI models, cannot achieve this goal. However, the additional complexity of QSMI models is enough to allow a model which recapitulates the outcomes of pair and trio growth experiments from Friedman et al.\cite{gore2017} with a single parameter set. We conclude that SMI models show the most promise in building models of larger microbial communities.

\subsection{Reversal of qualitative effects in general SSI models}
One consequence of SSI models, regardless of the specific choices of interaction functions $h_{ij}$, is that they imply a classification of the interaction between two microbes. That is, the sign of $h_{ij}(x_i,x_j)$ indicates if microbe $j$ has a positive effect on microbe $i$ ($h_{ij}(x_i,x_j) >0$), or a negative effect on microbe $i$ ($h_{ij}(x_i,x_j) <0$); recall that $h_{ij}(x_i,x_j)$ does not change sign for non-negative $x_i,x_j$. Generally, relationships are classified using both $h_{ij}(x_i,x_j)$ and $h_{ji}(x_j,x_i)$\cite{dohlman2019mapping}. SSI models allow us in some cases to predict the sign of the combined effects of two microbes on a third. Precisely, if two microbes have the same qualitative effect on the growth of a third, SSI models imply that their combined effect should be qualitatively the same (although of course different in magnitude). For example, if $h_{ij}(x_i,x_j)>0$ and $h_{ik}(x_i,x_k)>0$, then clearly $h_{ij}(x_i,x_j) + h_{ik}(x_i,x_k) >0$ for all non-negative $x_i,x_j,x_k$, and we should observe increased growth in microbe $i$ when in a trio with these two other species as compared to when grown alone. 

\begin{figure*}
\begin{subfigure}{0.45\textwidth}
\centering
    \includegraphics[scale = 1.5]{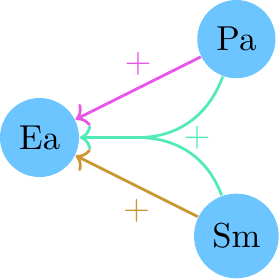}
    \caption{}
    \end{subfigure}
\begin{subfigure}{0.45\textwidth}
\centering
    \hspace{-1.5in}\includegraphics[scale = 1.5]{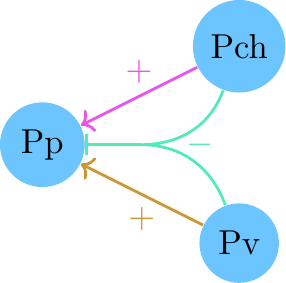}
    \caption{}
    \end{subfigure}
    \begin{subfigure}{0.45\textwidth}
\centering
    \includegraphics[scale = 0.43]{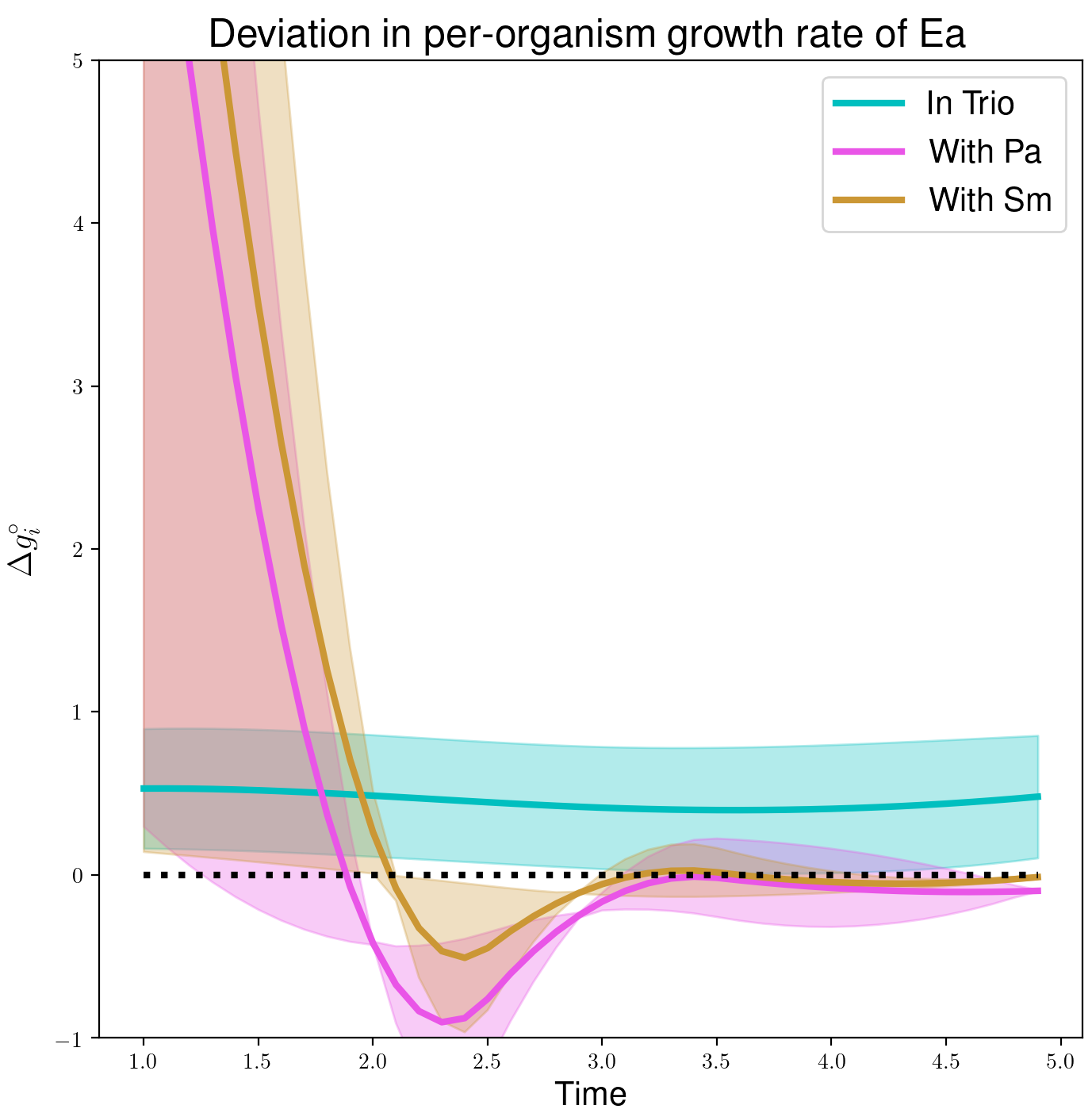}
    \caption{}
    \end{subfigure}
\begin{subfigure}{0.45\textwidth}
\centering
    \includegraphics[scale = 0.43]{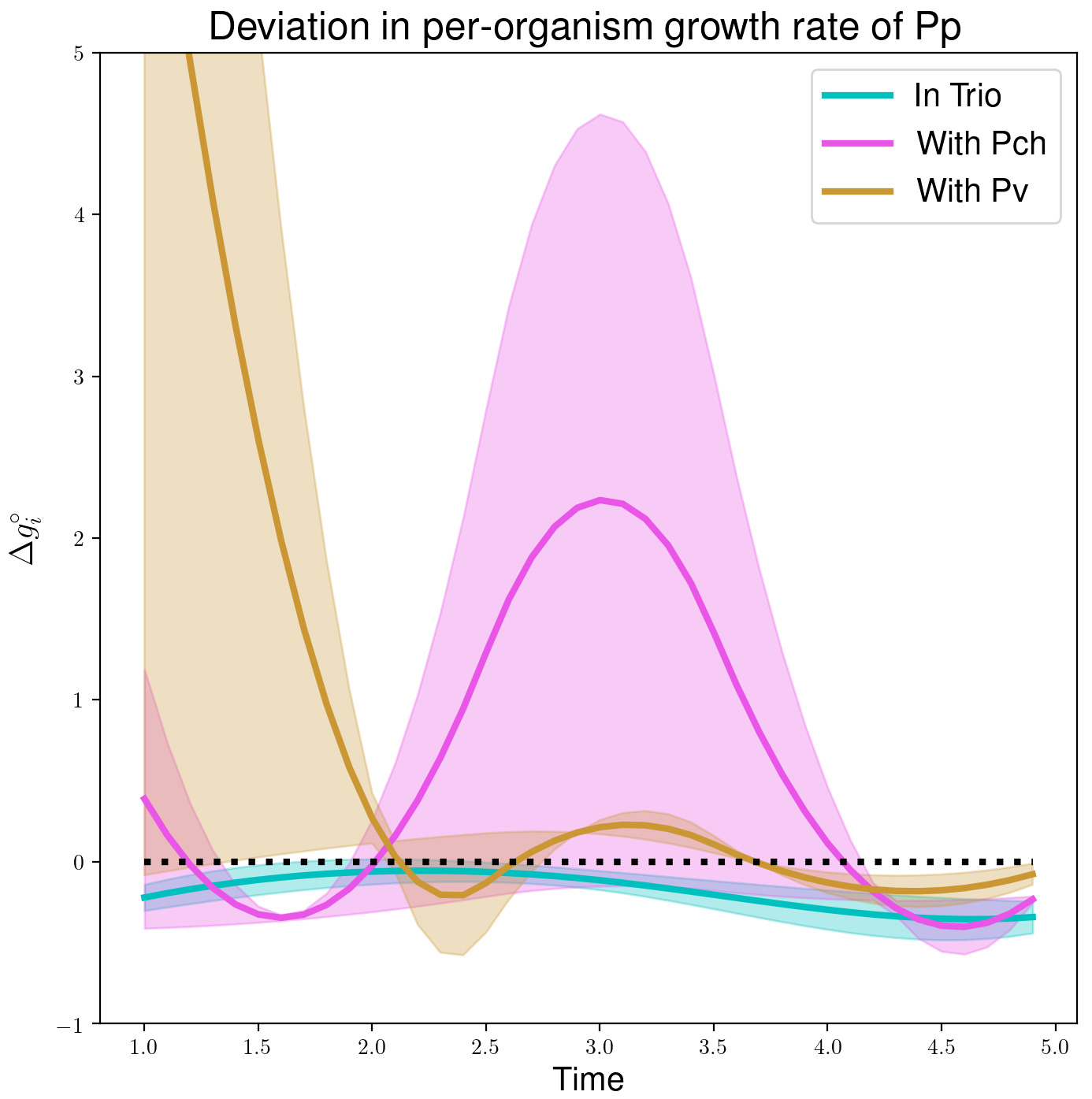}
    \caption{}
    \end{subfigure}
    \caption{(a,b) Arrows indicate influence on per-organism growth rate, as determined by deviation from growth rate when grown alone. Combined arrows represent the implied sum of influences. (c,d) Deviation in per-organism growth, estimated from time course data reported in Friedman et al.\cite{gore2017} (dashed line showing no deviation) (a,c) Pa (Pseudomonas aurantiaca, ATCC\#33663) and Sm (Serratia marcescens, ATCC\#13880) both have positive effects on the growth of Ea (Enterobacter aerogenes, ATCC\#13048) when grown in a pair. When these three are grown in a trio, the growth of Ea is promoted relative to its growth when grown alone. This trio makes sense in the context of SSI models. (b,d)Pch (Pseudomonas chlororaphis, ATCC\#9446) and Pv (Pseudomonas veronii, ATCC\#700474) both have positive effects on the growth of Pp (Pseudomonas putida, ATCC\#12633) when grown in a pair. However, when these three are grown in a trio, the growth of Pp is reduced compared to its growth when grown alone. This trio does not make sense in the context of SSI models.}\label{revers}
\end{figure*}

We estimate the effects on growth of microbes in pairs and trios by using the time-course data from Friedman et al.\cite{gore2017} to directly estimate the per-organism growth rate term of \cref{genLV}. We can therefore compare the effect on the per-organism growth rate of organism $i$ of being grown with species $j$ or species $k$ to the effect of being grown with both $j$ and $k$. We expect that if $j$ and $k$ both increase per-organism growth rate of species $i$, then the combined effect of species $j$ and $k$ is to increase per-organism growth rate of $i$, as in \cref{revers} (a). However, we do not always see this. In fact, we find that $11$ of the $54$ trios have at least one reversal of effect. \Cref{revers} (b) shows an example in which two species have a positive effect on the growth of a third when grown in pairs, but when grown in a trio there is reduced growth in the third species. All of the implied pairwise relationships are shown in \cref{networks} (a). The list of all trios with pair and trio estimated effects can be found in the supplemental file \verb|qualitative.csv|, and the 11 reversals are also found in \verb|reversals.csv|.

\begin{figure*}
\begin{subfigure}{0.45\textwidth}
\centering
\includegraphics[scale = 0.25]{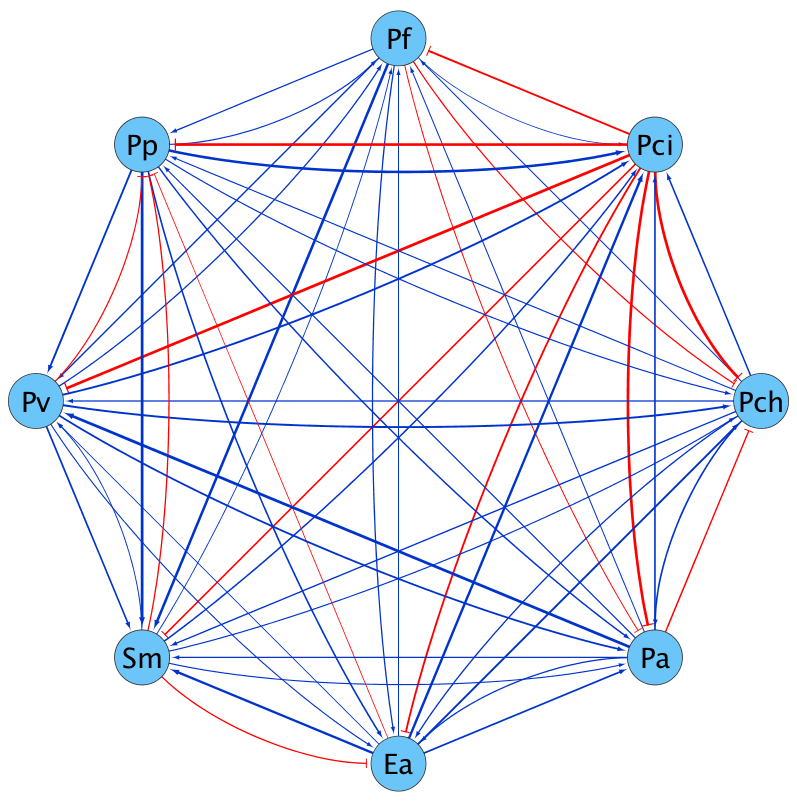}
\caption{}
\end{subfigure}
\begin{subfigure}{0.45\textwidth}
\centering
\includegraphics[scale = 0.25]{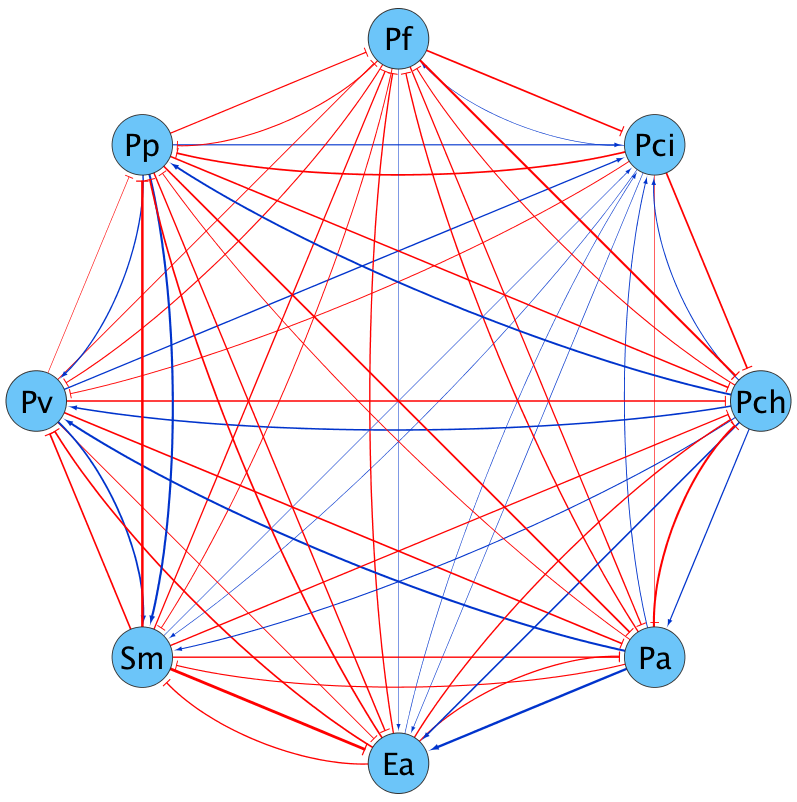}\caption{}
\end{subfigure}
\begin{subfigure}{0.45\textwidth}
\centering
\includegraphics[scale = 0.25]{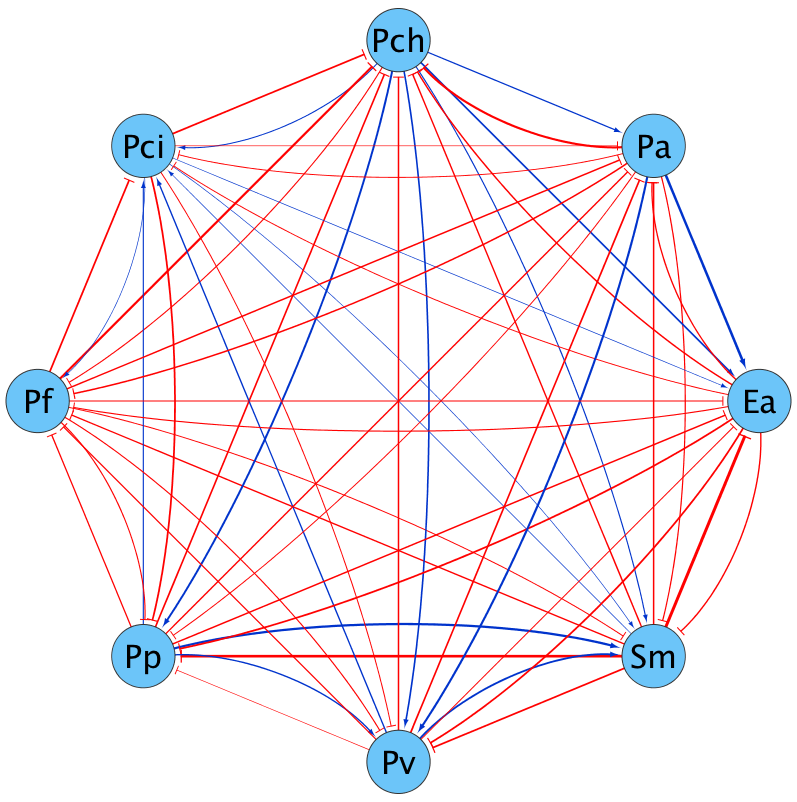}\caption{}
\end{subfigure}
\begin{subfigure}{0.45\textwidth}
\centering
\includegraphics[scale = 0.55]{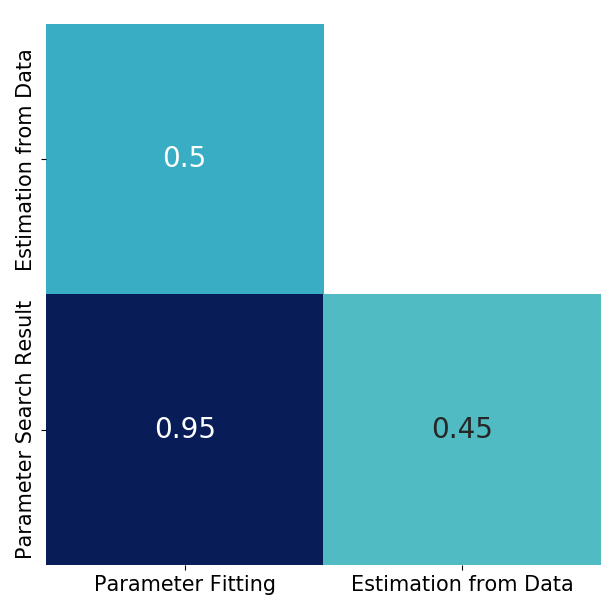}\caption{}
\end{subfigure}
    \caption{(a) Relationships between microbial species as implied by estimating average per-organism growth rate from growth data. (b)Relationships between microbial species as implied by parameters of the gLV model fitted to pair growth experiments. (c)Relationships between microbial species as implied by parameters of the gLV model from computational search for parameters to explain explain all experimental outcomes. (a,b,c)Blue arrows indicate positive influence on growth, red arrows indicate negative influence, and nodes are labeled with the species used, as abbreviated in Friedman et al.\cite{gore2017} \cite{gore2017} (Ea (Enterobacter aerogenes, ATCC\#13048), Pa (Pseudomonas aurantiaca, ATCC\#33663), Pch (Pseudomonas chlororaphis, ATCC\#9446), Pci (Pseudomonas citronellolis, ATCC\#13674), Pf (Pseudomonas fluorescens, ATCC\#13525), Pp (Pseudomonas putida, ATCC\#12633), Pv (Pseudomonas veronii, ATCC\#700474) and Sm (Serratia marcescens, ATCC\#13880)). (d) Proportion of edges in the network which match qualitatively (i.e. have the same sign) for each pair of networks from (a),(b), and (c).}
    \label{networks}
\end{figure*}

\subsection{Parameter fitting in the gLV model}

Above, we show that SSI models in general will not match the dynamics of the growth experiments from Friedman et al.\cite{gore2017}. However, it may be that SSI models, including the popular gLV model, give the correct qualitative outcomes in terms of survival and extinction over long time scales. In order to determine if this is the case, we fit a parameter set (the set of $\alpha_{ij}$ in \cref{LV}) to the time-course data of pair growth experiments, and ask if the model's asymptotic stability correctly predicts the outcome of the trio growth experiments. That is, we take as a model's ``prediction" the result of linear asymptotic stability analysis (see \cref{modout}).

We find that parameters fitted to pair growth experiments explain trio experiments for half of all trios, meaning that the model correctly predicts survival/extinction outcomes for half of all trios. Note that Friedman et al.\cite{gore2017} report that parameters fit to pairs lead to accurate predictions of 84\% of trios. However, this was calculated using a different definition of model prediction (see \cref{modout} for details). \Cref{networks}(b) shows the interactions implied by this parameter fitting. Interestingly, these interactions do not match the network of interactions determined using change in time-averaged growth, shown in \cref{networks}(a).

Additionally, we inspect long-time simulations of the gLV model for trios using parameters fitted to pair growth experiments, in order to determine if oscillatory behavior is predicted. We observe no oscillatory behavior in the simulations.

\begin{figure*}
	\begin{subfigure}{\textwidth}
		\begin{center}
			\includegraphics[scale= 0.5]{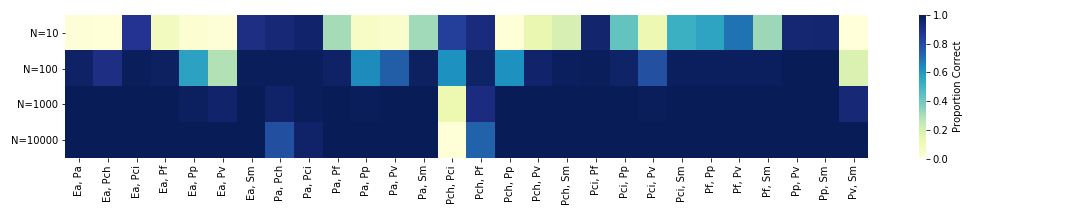}
		\end{center}
	\caption{Pair experiments with population size $10$, $100$, $1000$, $10000$.}
	\end{subfigure}
	\begin{subfigure}{\textwidth}
		\begin{center}
			\includegraphics[scale= 0.5]{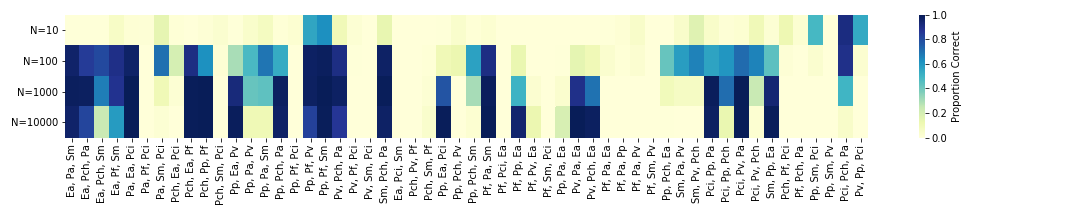}
		\end{center}
		\caption{Trio experiments with population size $10$, $100$, $1000$, $10000$.}
	\end{subfigure}
	\caption{Monte-Carlo experiments of the stochastic generalized Lotka-Volterra model, with parameters fitted to deterministic pair model. Darker color indicates that a larger proportion of the $1000$ trials used matched the observed growth experiment outcomes. Pairs and trios are abbreviated as in Friedman et al.\cite{gore2017}.}\label{montecar}
\end{figure*}

In order to determine if the model's failure to recapitulate experiments is the result of random fluctuations in growth, we use a stochastic version of the gLV model with the same parameter set. We determine the likelihood of the observed experimental outcome according to the stochastic gLV model, and see in \cref{montecar} that many outcomes have very low likelihood. Our experiment with the stochastic generalized Lotka-Volterra model shows that random fluctuation in growth is unlikely to explain the failure of the generalized Lotka-Volterra model to match the growth experiment data.

\subsection{The gLV model's capacity to recapitulate experimental outcomes}

We next determine if the gLV model has the capacity to recapitulate the outcome of the trio growth experiments while keeping a single parameter set that recapitulates pair experiment outcomes. We define the outcome of trio growth experiments in a binary fashion as reported by Friedman et al.\cite{gore2017}, and similarly define the outcome of pair growth experiments based on the final time-point of the experiment. 

Finding a single set of $\alpha_{ij}$ such that the model correctly predicts every pair and trio growth experiment would imply that the gLV model has the necessary complexity to match the qualitative outcomes of the growth experiments, if not the dynamics. However, we are unable to find this parameter set using a computational search with a pseudo-genetic algorithm (see \cref{search_meth}). Indeed, with the parameters resulting from this search, the model correctly predicts only 59\% of the trio outcomes. This suggests that the set of parameters for the gLV model which cause it to recapitulate the growth experiments is small or empty.

\begin{figure*}
	\begin{subfigure}{\textwidth}
		\includegraphics[scale=0.32]{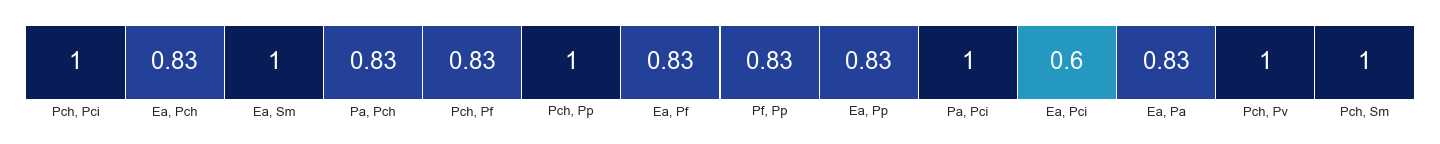}
	\end{subfigure}
	\begin{subfigure}{\textwidth}
		\includegraphics[scale=0.32]{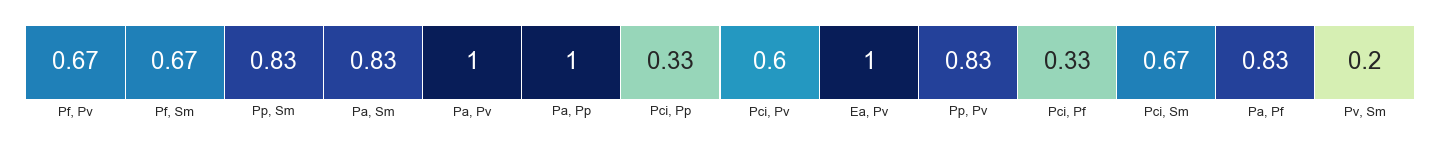}
	\end{subfigure}
	\caption{Proportion of explained trios as a result of each pair experiment, using the fitted interaction between that pair and varying the interactions involving the third member of a trio. For some pairs, parameters that explained the data are found for any third microbe, while for some parameters cannot be found for most third microbes. Pairs and trios are abbreviated as in Friedman et al.\cite{gore2017}.}\label{search}
\end{figure*}

To reduce the search space, we divide the growth experiments and attempt to find a parameter set for each group that predicts all of the qualitative outcomes of the growth experiments in that group. We use as groups all of the experiments involving one or both of some pair of microbial species (note that these groups are overlapping, but treated independently). For only $9$ of the $28$ pairs, parameters could be found that explained each trio involving that pair. \Cref{search} shows the proportion of trios involving a given pair which can be correctly predicted.

\subsection{A QSMI model to recapitulate growth experiments}

\begin{figure*}
    \centering
    \includegraphics[scale = 0.3]{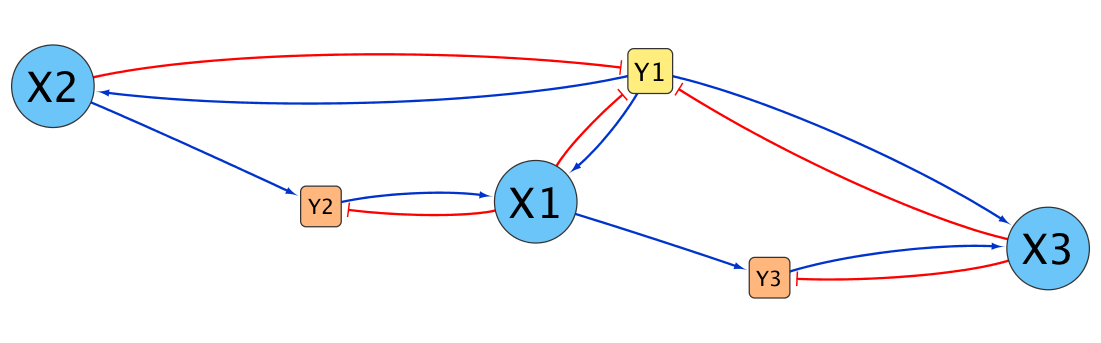}
    \caption{Network representation of a cross-feeding chain in which metabolism of $y_1$ by $x_2$ produces a byproduct which $x_1$ can metabolize for growth, and this process in turn produces a byproduct which $x_3$ can metabolize for growth. Note that the production of $y_2$ by $x_2$ only occurs is $y_1$ is present, and the production of $y_3$ by $x_1$ only occurs when $y_2$ is present, so that this chain does not have an effect on other trios involving a subset of three species in the chain. For simplicity, this dependency is not shown, nor is it included in \cref{metmodelnet}}
    \label{crossfeed}
\end{figure*}

 We ask if there exists a quadratic species-metabolite model which can match the qualitative outcomes of the growth experiments presented in Friedman et al.\cite{gore2017}. As with the generalized Lotka-Volterra model, we seek a single parameter set that can explain the interdependent growth experiments. We can build such a model by assuming the existence of one initially present metabolite along with additional molecules produced by the microbiota. These additional molecules allow us to form cross-talk chains, as shown in \cref{crossfeed}. This model also demonstrates that the situation detailed by \cref{revers}(b) can be modeled by a QSMI, using the mechanism of \cref{crossfeed} with $y_3$ acting instead as a poison to reduce growth of $x_3$.
 
For the pair experiments of Friedman et al.\cite{gore2017}, we need to add 19 cross-feeding molecules, bringing the total metabolites to 20. The pair models lead to trios for which we need to add cross-feeding chains to prevent a single extinction for 4 trios and prevent two extinctions for 1 trio. We also must implement cross-poisoning to cause a single extinction for 19 trios. For 1 trio, we need to adjust the model to cause one extinction and prevent another.

In total, we have a single model of 8 microbes and 72 molecules which are part of part of 19 pair-specific and 25 trio-specific cross-talk pathways. This model, when restricted by initial state to only two or three microbial species, recapitulates the outcomes of the growth experiments. 

\begin{figure*}
\centering
\includegraphics[scale = 0.6]{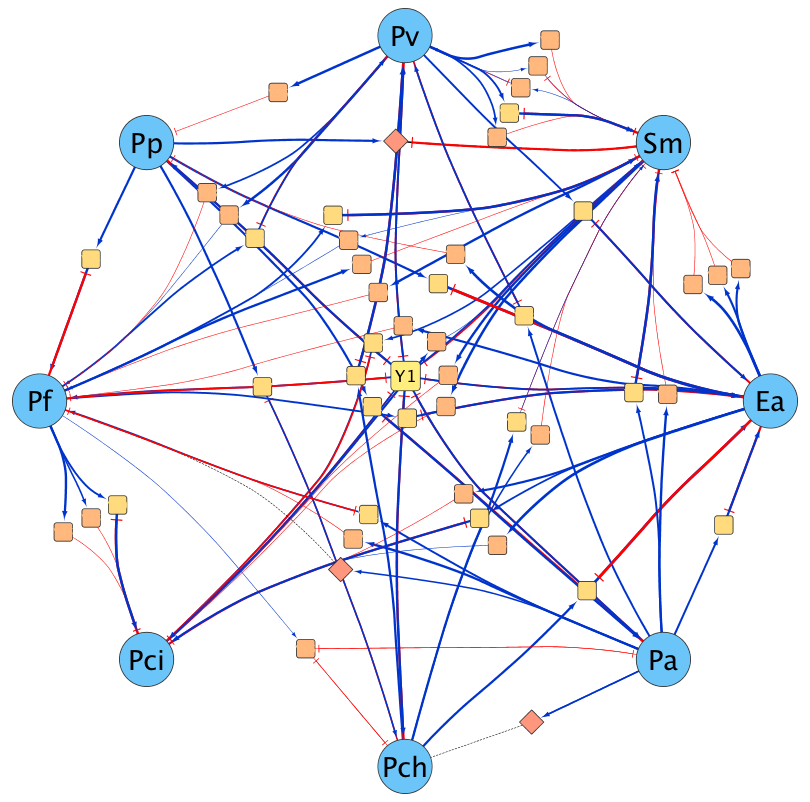}
\caption{Schematic of the full QSMI model that recapitulates growth experimental outcomes. This model contains the 8 microbes grown in Friedman et al.\cite{gore2017}, as well as 69 metabolites (squares), and 3 signalling molecules (diamonds). Blue arrows from metabolite to microbe indicate that the microbe uses that metabolite to grow, while red arrows from metabolite to microbe indicate that the metabolite inhibits growth of the microbe. Blue arrows from microbe to metabolite indicate production, and red arrows from microbe to metabolite indicate consumption or degradation. To match qualitative outcomes of growth experiments in Friedman et al.\cite{gore2017}, the model is run with only Y1 initially present, and initial microbial populations according to the growth experiment.}\label{metmodelnet}
\end{figure*}

\Cref{metmodelnet} shows the network of metabolite mediated interactions of the QSMI model. In this model, every microbe grows on a single metabolite, labeled $y_1$, and various pair or trio cross-talk chains alter that growth by providing extra resources (cross-feeding) or inhibiting a microbe's growth (cross-poisoning).

\subsection{Complexity of the QSMI model that recapitulates trio experiments.}

We wish to estimate the complexity of a QSMI that explains a given set of growth experiments. To do this, we reduce our set of allowable QSMI models to those that only include direct cross-talk as well as simple cross-talk chains such as that shown in \cref{crossfeed}, and detailed in \cref{y1star,y2star}. This restriction allows us to automate construction of a model that explains all but 2 of the trio experiments from Friedman et al.\cite{gore2017} using only 17 metabolites (this network can be viewed in \verb|min_met_network.csv| in the supplemental repository).

\begin{figure*}
    \centering
    \includegraphics[scale = 0.5]{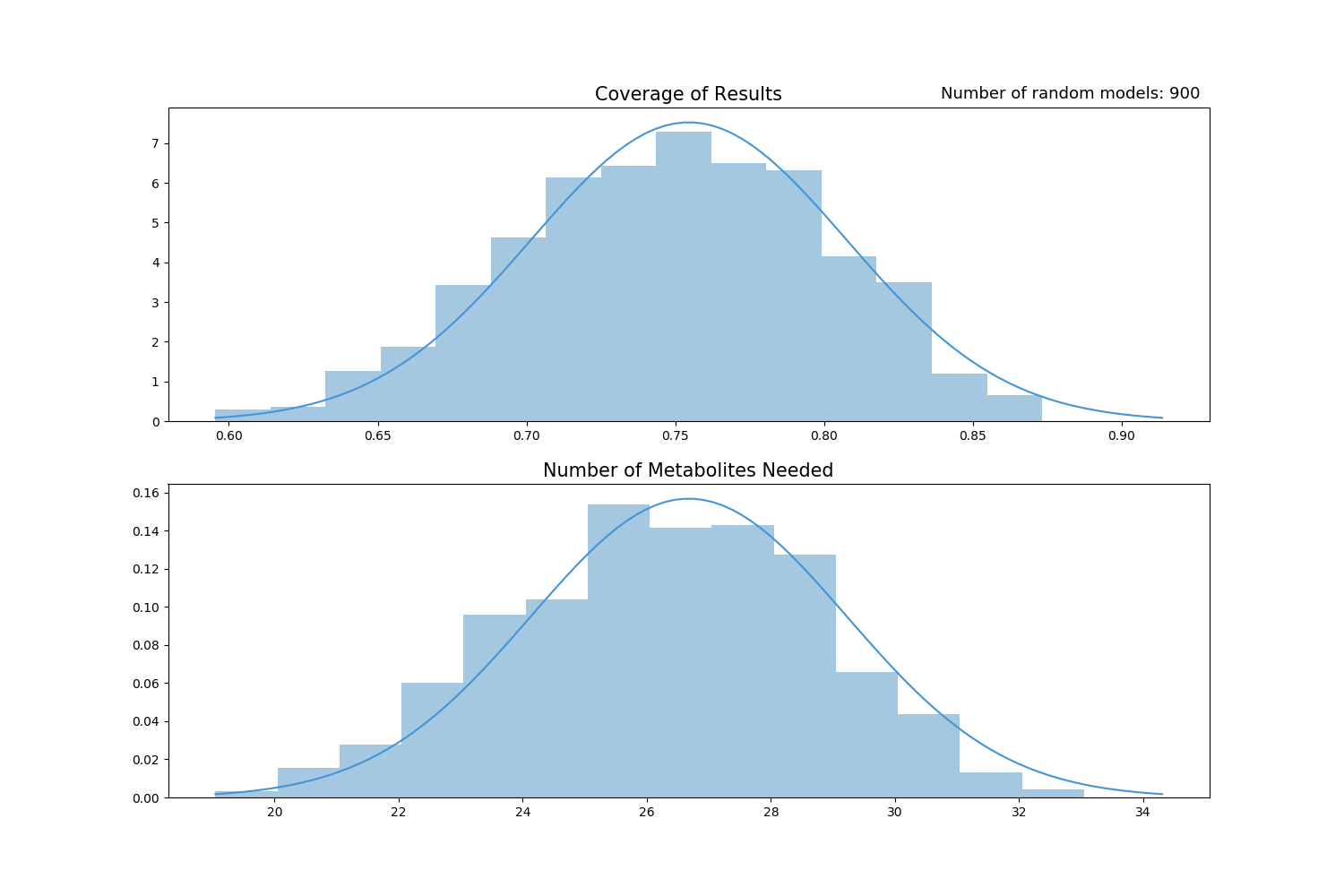}
    \caption{Histogram of coverage of randomized experimental outcome and number of necessary metabolites in an automatically generated QSMI model. Models are built to recapitulate randomly generated experimental growth outcomes and maximize prediction accuracy while minimizing number of metabolites.}
    \label{fig:random_models}
\end{figure*}

We next estimate the required complexity of a general QSMI model that recapitulates experiments. For a randomly generated set of growth outcomes, we determine how well an automatically generated QSMI model can recapitulate these outcomes, and how complex such a model must be. We hope that the best model generated for a given outcome set is of reasonable accuracy, explaining most or all of the data set, and complexity, requiring not too many metabolites. \Cref{fig:random_models} shows a histogram of the coverage of (i.e. what proportion of a set of outcomes is recapitulated) and number of metabolites needed in the best model we generate for a sample of randomly generated ``experimental" outcomes. The coverage achieved follows roughly a normal distribution with mean $0.754$ and standard deviation $0.053$, while the number of metabolites also follows roughly a normal distribution with mean $26.67$ and standard deviation $2.54$. This experiment demonstrates that QSMI models have the power to explain the majority of outcomes in a data set with a reasonable number of metabolites, and without pathways more complex than the one shown in \cref{crossfeed}. 

%
{\small
\section{Methods}\label{methods}
\subsection{Growth data used}
We use data from growth experiments published in Friedman et al.\cite{gore2017}. These experiments were carried out in flat-bottomed plates and grown in 48 hour growth-dilution cycles. Cell density was assessed using optical density (OD), and relative abundance was assessed by plating and colony counting. For this manuscript, we use units of OD computed by multiplying total culture OD by fraction of each species. Growth data is available in the supplemental material folder \verb|friedman_et_al_data|. For single microbe and pair growth experiments, Friedman et al.\cite{gore2017} provided intermediate time-course data. However, for trio growth experiments, only beginning and endpoints were reported, and so used here.


\subsection{Defining model prediction}\label{modout}
We define model prediction by the behavior of the model as time approaches infinity for any positive initial conditions. We compare this model prediction to experimentally observed extinction and coexistence, as reported in Friedman et al.\cite{gore2017}.

All models that we consider are ordinary differential equations (ODEs), with the exception of a stochastic analogue to an ODE model. For an ODE model, we define a model's outcome from the asymptotic stability of equilibrium, using standard methods (see for example \cite{ekesh}). That is, we write any ODE in the form
\begin{equation}\label{generalode}
    \frac{d\b{x}}{dt} = \b{f}(\b{x})
\end{equation}
where $\b{x}$ and $\b{f}$ may be vector-valued, and consider a point in the phase-space $\b{x}^*$ the ``outcome" or ``prediction" of the model if
\begin{equation}
    \b{f}(\b{x}^*) = 0
\end{equation}
and some solutions to \cref{generalode} approach $\b{x}^*$ and $t\rightarrow \infty$. Note that it is possible for there to exist more than one such $\b{x}^*$ for a single model (and parameter set), a condition known as \emph{bi-stability}\cite{ekesh}. In this case, we consider all such $\b{x}^*$ to be model outcomes.

For details on this process, see \cref{ASA}.

In Friedman et al.\cite{gore2017}, model outcome is defined using direct simulation rather than asymptotic stability.

\subsection{Qualitative effect on growth}

We use pair growth experiments to determine the qualitative effect of one species on another. That is, we find the difference between the time-average per-organism growth rate of species $i$ alone and in various pairs or trios.

We label the difference between per-organism growth rate in species $i$ when grown in some set of species and per-organism growth rate in species $i$ when grown alone as
\begin{multline*}
\Delta g_i^{\circ} = \textit{mean}\left(\left\{\frac{x^{\textit{with}( \circ)}(t+\Delta t)-x^{\textit{with} (\circ)}(t)}{x^{\textit{with}( \circ)}(t)}\right\}\right) \\- \textit{mean}\left(\left\{\frac{x^{\textit{alone}}(t+\Delta t )-x^{\textit{alone}}(t)}{x^{\textit{alone}}(t)}\right\}\right)
\end{multline*}
where, for example $\circ$ is $j$ for species $i$ grown a pair with species $i$, and $\cdot$ is $j,k$ for species $i$ is grown in a trio with $j$ and $k$. Then, we observe that the additivity of \cref{genLV} and the assumption that the functions $h_{ij},h_{ik}$ do not switch sign imply that, if the microbiota grows according to \cref{genLV}, then
\begin{equation}
    \Delta g_i^j > 0 \& \Delta g_i^k > 0 \Rightarrow \Delta g_{i}^{jk} >0
\end{equation}
and 
\begin{equation}
    \Delta g_i^j < 0 \& \Delta g_i^k < 0 \Rightarrow \Delta g_{i}^{jk} <0.
\end{equation}
In other words, if two species have the same qualitative effect on the growth of a third, their combination will have that same qualitative effect. We simply compare these quantities in the data to find examples for which this does not hold. We call such an instance a ``reversal". Note that the serial dilution of the growth experiment mean that a reversal of effect is not the result of crowding.

\subsection{Generalized Lotka-Volterra parameter fitting}\label{search_meth}

We use non-linear least squares procedures to fit parameters of the gLV model to the time-course experimental data. Additionally, we use a pseudo-genetic algorithm to attempt to find a single set of parameters $\{\alpha_{ij}\}$ which cause the model to recapitulate the growth experiment outcomes.

We fit a set of parameters from \cref{LV} to the time-course data. For individual growth parameters ($r_i,K_i$ in \cref{LV}), we use individual growth data, and non-linear least squares (implemented in python package scipy.optimize\cite{scipy}) to fit a logistic curve. We fit the parameters $\alpha_{ij}$ to pair growth experiment data. We again using non-linear least squares (implemented in python package scipy.optimize\cite{scipy}), computing solution curves for a given parameter set numerically in order to compute residuals. The parameters fitted, along with code to reproduce the fitting, can be found in the supplemental material. 

We also ask whether or not the model has the capacity to explain all of the growth experiment outcomes with a single parameter set by searching for such a parameter set. Notice that the interdependence of the trios mean that this is a stricter condition than the existence of a parameter set for each trio which correctly predicts the experimental outcome of just that trio, and also stricter than the existence of a single such parameter set for each trio and involved pairs.

We take advantage of qualitative analysis, detailed in Appendix A, to search for parameter values that explain all of the outcomes observed. To perform this search, we begin with parameter sets that explain 12 independent trios. We then use a pseudo-genetic algorithm, assessing the fitness of the parameter set by the magnitude of the eigenvalues of the various Jacobian matrices whose sign do not match the eigenvalues that would result in a model matching the observed data, to search for a parameter set which allows the model to recapitulate the experiments. In this algorithm, genes can be mutated with a continuous random variable, whereas in a standard genetic algorithm genes are generally taken over a discrete set \cite{mccall2005}. See Appendix A and supplementary materials for explanation and code, respectively. The parameters found can be found in the supplemental material.

We next relax the search condition by attempting to find for each pair of species parameters that gave an accurate prediction of all the trios they are involved in without changing the interaction parameters between that pair. This is done in order to identify pairwise relationships that are consistent across different trios. To do this for a pair of species $i,j$, we first fix $\alpha_{ij}$ and $\alpha_{ji}$ as fitted to time-course data. Then, for eack other species $k$, we seek $\alpha_{ik}$, $\alpha_{jk}$, $\alpha_{ki}$, $\alpha_{kj}$ so that model asymptotic stability matches the observed outcome of the experiments. We again use a pseudo-genetic algorithm, taking advantage of some partial conditions for equilibrium stability given in the supplementary code. We record for each pair the proportion of trios for which parameters can be found to explain the experimental outcomes. The result is shown in \cref{search}.

\subsection{The stochastic generalized Lotka-Volterra model}

We choose the stochastic process which has the property that as we increase the concentration of organisms modeled we recover the deterministic generalized Lotka-Volterra model\cite{kurtz72}. This property means that rather than simply ``adding noise" to the model, we have chosen the closest fully stochastic analogue to the model, which is also the standard choice of stochastic model for interactions between agents under the assumption of mass action kinetics \cite{anderson2011continuous,anderson2015,anderson_kurtz}. See \cref{stoch} for details of the model.

We produce (exact) realizations of the stochastic process using Gillespie's algorithm\cite{gillespie1976} (sometimes called the Stochastic Simulation Algorithm), as shown in \cref{stoch_reali}. Additionally, we perform Monte Carlo experiments to estimate the probability ($\pm 6 \%$) of each observed outcome according to the stochastic model with parameters fitted to pairs. We use the $\tau$-leaping algorithm from \cite{anderson2007} for efficiency.

\begin{figure*}
	\begin{subfigure}{0.48\textwidth}
		\includegraphics[scale= 0.24]{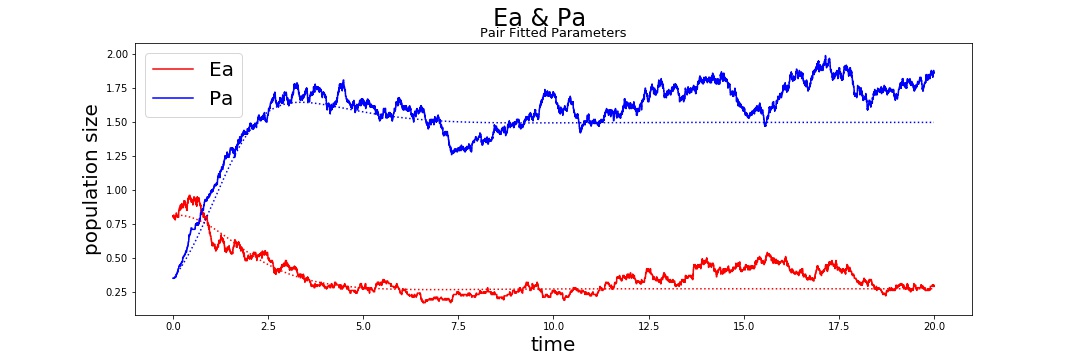}
		\caption{}
	\end{subfigure}
	\begin{subfigure}{0.48\textwidth}
		\includegraphics[scale= 0.24]{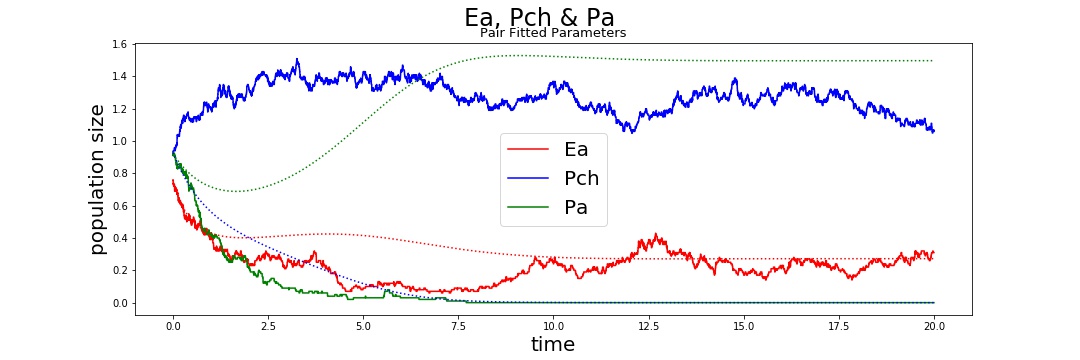}
		\caption{}
	\end{subfigure}
	\caption{Exact realizations of stochastic generalized Lotka-Volterra, with parameters fitted to deterministic pair model for species Ea, Pa, and Pch (as abbreviated in Friedman et al.\cite{gore2017}). Notice that in realization (b), we have an extinction event that is not implied by the deterministic model (shown dashed).}\label{stoch_reali}
\end{figure*}

\subsection{Construction of QSMI Model}

Starting with parameters given by models of individual growth, we can arrive at every possible pair outcome by assuming there are at most $\binom{n}{2}+1$ external metabolites used for growth, but only one is initially present. We build this model by adding cross-talk molecules to models of growth on a single nutrient. In our model, these cross-talk molecules are produced by one microbe and have some effect on another. 

We do this systematically by adjusting pair or trio models separately. This means that we add a cross-talk pathway to a pair model so that it recapitulates the correct growth experiment outcome, and does so in such a way that the added cross-talk pathway will have no effect on any other pair, and likewise with trios. That is, any cross-talk pathway must require all members of the pair or trio it was added for to be present to have some effect. This restriction makes the construction systematic, but is not necessary.

We begin with a model of growth:
\begin{align}
    \frac{dx_1}{dt} &= \ka_{11}x_1y_1 - d_1 x_1\\
    \frac{dy_1}{dt} &= f_1 - d^*_1y_1 -\ka_{11}x_1 y_1.
\end{align}
For each pair model, if the initial model formed simply by including both microbes does not match growth outcome, we add a metabolite to the model. Restricted to a pair, the model is then 
\begin{align}
        \frac{dx_1}{dt} &= \ka_{11}x_1y_1 - d_1 x_1 + \psi_{12}x_1y_2 \label{xtalk1}\\
        \frac{dx_2}{dt} &= \ka_{21}x_2y_1 - d_2 x_2  \label{xtalk2}\\
        \frac{dy_1}{dt} &= f_1 - d^*_1y_1 -\ka_{11}x_1 y_1- \ka_{21}x_2 y_1  \label{xtalk3}\\
        \frac{dy_2}{dt} &= \ka_{21}x_2 y_1 - d^*_2 y_2 -\ka_{12} x_1y_2  \label{xtalk4}
\end{align}
where $y_2$ serves as the cross-talk molecule, in this example produced by $x_2$ and having some effect (determined by $\psi_{12}$) on $x_1$. 

To adjust the model to account for trio growth experiments, we add trio-specific cross-talk pathways. Precisely, given a model of organisms $x_1,x_2,x_3$ (assuming without loss of generality that $x_1$ survives) which may already include pair-wise cross-talk and cross-poisoning, we can introduce a trio specific chain by introducing two new metabolites $y_1^*,y^*_2$ such that
\begin{align}
    \frac{dy_1^*}{dt} &= \ka_{11} x_1 y_1 - d^{**}_1 y_1^* -\ka_{21}^* x_2y_1^* \label{y1star}\\
    \frac{dy_2^*}{dt} &= \ka_{21}^* x_2 y_1^* - d^{**}_2 y_2^* -\ka_{32}^* x_3y_2^*\label{y2star}
\end{align}
where $y_2^*$ has some effect on $x_3$. These equations model a situation in which $y_1^*$ is produced as a metabolic by-product of $x_1$ metabolizing $y_1$, and likewise $y_2^*$ is produced as a metabolic by-product of $x_2$ metabolizing $y_1^*$. Then, $x_1$ and $x_2$ must both be present for $y_2^*$ to be produced. This chain is specific to this trio as long as $y_2^*$ only has an effect on $x_3$.

We must also for one trio change the model so that it predicts coexistence of the entire trio rather than two extinctions. This requires a signaling molecule and a positive feedback loop of cross-feeding among the two species which were previously predicted to go extinct. This model therefore needs three additional molecules.

Finally, there is one trio for which we need the model to predict one lone survivor instead of a different lone survivor. This requires a signal that is not consumed or diluted (or was diluted at a very small rate). To do this, we use two signalling molecules that are never degraded or consumed, and cause one species to poison the other two species with a third molecule. 

The complete model is detailed in the supplemental material, and a schematic of the interactions in the model is shown in \cref{metmodelnet}.

\subsection{Estimation of QSMI model complexity}

To estimate QSMI model complexity, we reduce our set of allowed models to only those with one initial available metabolite and direct cross-talk between pairs as well as cross-talk chains as detailed in \cref{y1star,y2star}. This allows us to automatically generate a model which explains a large proportion ($98\%$ in the case of the data from Friedman et al.\cite{gore2017} - all trios beside the last two detailed above) of a set of pair and trio growth experimental outcomes. We attempt to maximize this coverage and minimize the number of molecules needed in order to estimate the complexity of QSMI models. In our construction, the number of additional pathways added is ultimately a function of the microbes relative ability to metabolize the initial available metabolite. This observation allows us to optimize over the set of possible orders of these metabolic parameters. 

We generate random trio and growth experiment outcomes by permuting the real experimental outcomes. In this way, we preserve the number of each possible qualitative outcome (i.e. coexistence, extinction, and double extinction). We then generate the best possible model for each random outcome set.

}
%

\section{Discussion}

We would like to build a clinically useful model of the dynamics of the human microbiome. For this, we seek a modeling framework that infers community dynamics from fundamental interactions, so that data and discoveries from across studies can be incorporated into an individualized model using these interactions as building blocks. We therefore need a model that can be built without reparameterization and can capture emergent properties of microbial community composition dynamics. 

As a representative example of species-species modeling, we inspect the generalized Lotka-Volterra model. Using this model, we see disagreement between trio growth experiments and model prediction based on pairwise fitted parameters, and we are unable to find a set of parameters which allow the model to recapitulate the qualitative outcomes of the interdependent growth experiments. This suggests that the generalized Lotka-Volterra model has a high sensitivity to fitted parameters, even for qualitative results, and that the space of parameters which fit the entire set of qualitative growth experiments is small or does not exist. 

In species-metabolite modeling, dynamics are modeled by the interactions of individual microbes with a shared metabolite pool. We find that this framework has the additional complexity necessary for capturing emergent behavior through cross-feeding and cross-poisoning. We show that this framework does not adhere to the additive interaction assumption, and that a model can be found to fit interdependent growth data. We then use the mechanisms of cross-feeding and cross-poisoning to fit various competitive configurations to complex outcomes.


It is worth noting that SMI models are inherently more complex than SSI models, as proven by Momeni et al.\cite{momeni2017lotka}. Our result, which relates SSI and SMI modeling to experimental growth data, compliments the conclusion of Momeni et al.\cite{momeni2017lotka} by showing that the dynamical complexity is indeed necessary for accurate modeling. The appeal of SMI models, however, goes beyond mere complexity. SMI models may provide better fundamental building blocks for inferring community dynamics because they better reflect the real biological building blocks of microbial community interactions. This allows us to build models for larger communities by simply combining models for smaller communities. In contrast, one might add higher order terms to an SSI model to account for interactions between more than two species. However, this approach requires that a model built for a large community has little relationship with models of sub-communities. Such an extension of SSI models therefore does not achieve our goal of individualized modeling.



\medskip

Our work focuses on establishing the capacity of different modeling frameworks for recapitulating the emergent behavior of relatively simple microbial communities. While this work provides an important starting point, it is nonetheless limited in applicability. For instance, it is fair to say that the procedure to build the QSMI model for the growth experiments in Friedman et al.\cite{gore2017} is not in the spirit of building a model in an individualized manner. In practice, cross-feeding and other interactions of the microbe with the metabolite pool might be discovered from, for example, metabolic modeling, and SMI models may be built from genome-scale metabolic models of individual microbial species\cite{chan2017,Mendes-Soares2016,zomorrodi2014}. In this way, high-throughput sequencing technology can be leveraged to better understand microbiome composition. Developing methods to build clinically useful species-metabolite models with available data remains an open and interesting area of research.

In addition, we investigate the complexity of the QSMI model by minimizing the number of metabolites needed to match most of a set of growth experiment outcomes. It would also be interesting to build the QSMI model with further ``reality" criteria in mind, such as using the minimum number of cross-talk pathways. In future work, we plan to establish a systematic method to build a QSMI model which matches some set of outcomes exactly and satisfies various reality conditions. It is also of interest to establish easily check-able conditions on a set of growth experiment outcomes which decide whether or not there exists a QSMI which can recapitulate the experiments. While these questions are very interesting, they are fundamentally mathematical considerations and outside of the scope of this paper, which attempts to determine model usefulness in the context of real data. 

Species-metabolite interaction models provide an intermediate level of complexity between fully detailed genome-scale models and fully simplified models such as species-species interaction models. This level of complexity holds promise for individualized predictions in medicine.

\bigskip
{\it This work was supported by funding from the Andersen Family Foundation, National Cancer Institute grant R01 CA179243, and the Center for Individualized Medicine, Mayo Clinic.}

{\it Code for parameter estimation and searching, parameter values, and a complete description of the QSMI model, as well as code for the stochastic experiments, is available at \\\verb|https://github.com/jdbrunner/model_comparisons|}.

{\it JB carried out computational, statistical, and mathematical analysis, and drafted the manuscript. NC directed the goals of the analysis, and critically revised the manuscript. All authors gave final approval for publication and agree to be held accountable for the work performed therein.}

\bibliographystyle{siam}
\bibliography{subnet}

\begin{appendices}

\section{The Positive Steady State of the generalized Lotka-Volterra model}\label{SSofLV}

\subsection{Asymptotic stability analysis}\label{ASA}

We can re-scale \cref{LV} with two species to 
\begin{align}\label{LVrescaled}
\frac{d{x}_1}{dt} = r_1 x_1 (1-x_1 + \alpha_{12}x_2)\\
\frac{d{x}_2}{dt} = r_2x_2(1-x_2 + \alpha_{21}x_1)\nonumber
\end{align}
in order to simplify notation, and analyze asymptotic behavior of this model by performing straightforward stability analysis on the equilibrium\cite{ekesh}. We see that \cref{LVrescaled} has equilibrium at $(0,0)$, $(1,0)$, $(0,1)$ and 
\[
\b{x}^* = \left(\frac{1+\alpha_{12}}{1- \alpha_{12}\alpha_{21}},\frac{1 + \alpha_{21}}{1- \alpha_{12}\alpha_{21}}\right)
\]
Furthermore, linearization about each of those points reveals that $(0,0)$ is never stable, $(1,0)$ is stable if $\alpha_{21}<-1$, $(0,1)$ is stable if $\alpha_{12} < -1$. Lastly, $\b{x}^*\in \bR^2_{\geq 0}$ if and only if $\{\alpha_{12} <-1 ,\alpha_{21}<-1\}$ or $\{-1<\alpha_{12}, -1<\alpha_{21}, \alpha_{12}\alpha_{21}< 1\}$ and if $\{\alpha_{12}<-1, \alpha_{21} < -1\}$, then $\b{x}^*$ is unstable, and is in fact a saddle point. If $\{-1<\alpha_{12}, -1<\alpha_{21}, \alpha_{12}\alpha_{21}< 1\}$, then $\b{x}^*$ is stable. All of this can be done through symbolic analysis of the Jacobian matrix evaluated at $\b{x}^*$.

We can now characterize the outcomes observed in the paper using the parameters $\alpha_{12}$ and $\alpha_{21}$:
\begin{enumerate}[(a)]
	\item Coexistence: this is stability of the positive state, and so requires $\{-1<\alpha_{12}, -1<\alpha_{21}, \alpha_{12}\alpha_{21}< 1\}$.
	\item Invasion of one species regardless of initial condition: this is stability of one boundary state \emph{and instability of the other}. This requires $\alpha_{12} <-1$, $\alpha_{21} > -1$ or the opposite. If $\alpha_{12}<-1$ then 2 invades 1.
	\item Bi-stability: This is stability of both boundary states, and requires $\{\alpha_{12}<-1, \alpha_{21} < -1\}$.
\end{enumerate}
Interestingly, case (c) is not observed in the data of \cite{gore2017}. 

The three species model is
\begin{align}
\frac{d{x}_1}{dt} = r_1 x_1 (1-x_1 + \alpha_{12}x_2 + \alpha_{13}x_3)\\
\frac{d{x}_2}{dt} = r_2x_2(1-x_2 + \alpha_{21}x_1 + \alpha_{23}x_3)\\
\frac{d{x}_3}{dt} = r_3x_3(1-x_3 + \alpha_{31}x_1 + \alpha_{32}x_2)
\end{align}
and here again we can compute model equilibrium states and stability. There are 8 equilibrium points, corresponding to each qualitative possibility of survival \& extinction. Again, the $(0,0,0)$ equilibrium is never stable. There exist simple conditions on the parameters for local stability of all equilibrium points except for the state which represents coexistence of all three microbes. Stability of this last state can, however, be easily evaluated for any given parameters. 

We can compute the Jacobian determinant to see that the stability conditions for the double extinction equilibrium points are
\begin{itemize}
	\item $(1,0,0):\;\alpha_{21} < -1,\, \alpha_{31}<-1$
	\item $(0,1,0):\;\alpha_{12}<-1,\,\alpha_{32}<-1$
	\item $(0,0,1):\;\alpha_{13}<-1,\,\alpha_{23}<-1$
\end{itemize}

Taking advantage of symmetry, we investigate only one of the three single extinction equilibrium, which have the form $\left(\frac{1+\alpha_{12}}{1- \alpha_{12}\alpha_{21}},\frac{1+\alpha_{21}}{1 - \alpha_{12}\alpha_{21}},0\right)$. The first two eigenvalues of the Jacobian matrix at these points will follow the two dimensional case, so we have the necessary conditions for stability $\{-1<\alpha_{12}, -1<\alpha_{21}, \alpha_{12}\alpha_{21}< 1\}$. This is simply because after extinction of species $k$, the model is identical to the pair model. While unsurprising, this fact does imply that not all hypothetical combinations of existence and extinction outcomes for pair and trio experiments can be simultaneously explained by the parameters of the generalized Lotka-Volterra model. However, there were no instances in the trio experiments being considered in which such a ``smoking gun" scenario was observed.

The third eigenvalue is the value of $r_3(1-2x_3 + \alpha_{31}x_1 + \alpha_{32}x_2)$ evaluated at this point, which is
\begin{equation}
\lambda_3 = r_3\left(1+ \alpha_{31}\frac{1+\alpha_{12}}{1- \alpha_{12}\alpha_{21}} + \alpha_{32}\frac{1+\alpha_{21}}{1 - \alpha_{12}\alpha_{21}}\right)
\end{equation}
Clearly if $\alpha_{31}$ and $\alpha_{32}$ are both positive, this state is unstable. The condition for linear stability is
\begin{equation}\label{sing_ext}
\alpha_{31}\frac{1+\alpha_{12}}{1- \alpha_{12}\alpha_{21}} + \alpha_{32}\frac{1+\alpha_{21}}{1 - \alpha_{12}\alpha_{21}} < -1.
\end{equation}

\subsection{Lack of limit cycles of the two-species gLV model}

We can rule out closed orbits in the two species gLV model using Dulac's criterion. Letting
\begin{equation}
    g(x_1,x_2) = \frac{1}{x_1 x_2}
\end{equation}
we compute
\begin{equation}
    \nabla \cdot \left((\frac{d{x}_1}{dt},\frac{d{x}_2}{dt})g(x_1,x_2)\right) = -\frac{r_1}{x_2}-\frac{r_2}{x_1} < 0 
\end{equation}
for all $x_1,x_2 > 0$. This implies that there are no solution to \cref{LVrescaled} is a closed orbit in the positive quadrant.

Note the that the standard predator-prey ``Lotka-Volterra" model does allow closed orbits. This is because that model does not include the quadratic terms $-r_1x_1^2$ and $-r_2x_2^2$ that appear in \cref{LVrescaled}, and because that model does not enforce the assumption that $r_i >0$. This can be interpreted as an assumption of infinite carrying capacity of the prey species and exponential decay of the predator.

\subsection{Pseudo-Genetic Algorithm}

We search for a parameter set to fit qualitative growth behavior by performing a pseudo-genetic algorithm which attempts to minimize 
\begin{equation}
    M = \sum_{trios}\left(\sum_{\Lambda_i} \lambda^* + p_i\right)
\end{equation}
where $\Lambda_i$ is the set of eigenvalues which corresponded to the equilibrium point which matches the experimental outcome of trio $i$, and $p_i = 0$ if the three pairs of trio $i$ match experimental outcome, and $p_i = 1000$ otherwise. The chromosomes of the genetic search are taken to be the parameter sets, represented as a matrix whose $i,j$ entry contained $\alpha_{ij}$. We use the rows of this matrix as genes, and so the mating procedure is to choose for each row of the child the row of one or the other parent with even probability.

We describe this as a ``pseudo-genetic" algorithm because we are searching over a continuous parameter space. In order to account for this, random mutation of parameters is done by perturbation with a continuous random variable. First, to determine if mutation occurred, we draw a uniform random variable in $(0,1)$ and mutate if this variable is less than a thresh-hold of $0.2$. If mutation occurs, a random matrix whose entries are generated uniformly in $[-0.05,0.05]$ is added to the matrix representing the parameter set.

\subsection{The stochastic generalized Lotka-Volterra model}\label{stoch}
The model is as follows: 

\begin{multline}	
X_i(t) = X_i(0) + Y^1_i\left(r_i\int_{0}^t X_i(s)ds\right) \\- Y_i^2\left(\hat{r}_i\int_0^t X_i(s)(X_i(s)-1)ds\right)\\ + \sum_{i\neq j}Y_{ij}\left(\hat{\alpha}_{ij}\int_0^t X_i(s)X_j(s)ds\right).
\end{multline}

Here, $Y(p(t))$ are non-homogeneous Poisson (jump) processes with time-varying propensity $p(t) = \int_0^t f(s)ds$. The new parameters $\hat{r}_i$ and $\hat{\alpha}_{ij}$ depend on the ``volume" of the experiment, i.e. the population size scale. Precisely, with a volume $N$ we take $\alpha_{ij}$ as fitted to pair growth experiments and let
\[
 \hat{r_i} = \frac{r_i}{N} \quad \hat{\alpha}_{ij} = \frac{\alpha_{ij}}{N}
\]
Then, as $N\rightarrow \infty$, realizations of the stochastic model $\frac{X}{N}$ approach trajectories of the deterministic model \cite{kurtz72}.

\section{Stability of equilibrium of QSMI model.}\label{stability}

Consider the model for $n$ microbes
\begin{align}
    \frac{d}{dt}x_i &= \ka_{i}x_iy - d_i x_i\qquad i = 1,..,n\\
    \frac{d}{dt}y &= f_{y}  - d_y y - \sum_{i=1}^n \ka_{i}x_i y 
\end{align}
This has equilibrium at $x_i=0, y = \frac{f_y}{d_y}$, and at $y=\frac{d_i}{\ka_i}$ for each $i$, with $x_j >0$ if and only if $\frac{d_j}{\ka_j} = \frac{d_i}{\ka_i}$. The general form of the characteristic equation of the Jacobian matrix about any steady state for this system is
\begin{equation}
    \det(J-\lambda I) = (-d_y -\sum_{i=1}^n\ka_i x_i - \lambda)\prod_{i=1}^n (\ka_i y -d_i - \lambda) + \sum_{i=1}^n \ka_i^2 y x_i \prod_{j\neq i} (\ka_j y - d_j -\lambda) 
\end{equation}

Solving at the extinction steady state $x_i=0, y = \frac{f_y}{d_y}$, we have eigenvalues $\lambda_{n+1} = -d_y$, $\lambda_j = \frac{\ka_j f_y}{d_y} - d_j$. Therefore, this state is linearly stable if and only if
\begin{equation}
    d_i d_y > k_i f_y \; \forall i 
\end{equation}

Next, for each $i$ let $\cQ_i = \left\{j|\frac{d_j}{\ka_j} = \frac{d_i}{\ka_i}\right\}$. Then for each $i$ we have the set of equilibrium defined by 
\begin{equation}
    \sum_{j\in \cQ_i}  \ka_j x_j = \frac{\ka_i}{d_i}\left(f_y - d_y \frac{d_i}{\ka_i}\right)
\end{equation}
and $x_l = 0$ if $l \not \in \cQ_i$. The characteristic equation becomes
\begin{equation}
    \det(J-\lambda I) = -\left[\lambda^2  + \lambda d_y + \lambda\frac{\ka_i}{d_i}\left(f_y - d_y \frac{d_i}{\ka_i}\right) + \sum_{j \in \cQ_i}\ka_j^2 x_j\right]\lambda^{|\cQ_i|-1}\prod_{l \not \in \cQ_i}\left(\ka_l \frac{d_i}{\ka_i}-d_l - \lambda\right)
\end{equation}
First, we see that if $\frac{d_i}{\ka_i} \neq \min_{l = 1,...,n} \left\{\frac{d_l}{\ka_l}\right\}$, then this is unstable. If we do have the minimum $\frac{d_i}{\ka_i}$, then the remaining nonzero eigenvalues are
\begin{equation}
    \lambda_{\pm} = \frac{1}{2}\left[-B \pm \left(B^2 + 4A\right)^{\nhalf}\right]
\end{equation}
where $B =  d_y + \frac{\ka_i}{d_i}\left(f_y - d_y \frac{d_i}{\ka_i}\right) >0$ and $A = \sum_{j \in \cQ_i}\ka_j^2 x_j>0$. These both then have negative real part, implying that the hyperplane of solutions is attracting (note that if $|\cQ_i| =1$, this implies a linearly stable equilibrium point). 

Next, we consider the two species cross-feeding or cross-poisoning model:
\begin{align}
        \frac{d}{dt}x_1 &= \ka_{11}x_1y_1 - d_1 x_1 + \psi_{12}x_1y_2 \label{xtalk1c}\\
        \frac{d}{dt}x_2 &= \ka_{21}x_2y_1 - d_2 x_2  \label{xtalk2c}\\
        \frac{d}{dt}y_1 &= f_1 - d^*_1y_1 -\ka_{11}x_1 y_1- \ka_{21}x_2 y_1  \label{xtalk3c}\\
        \frac{d}{dt}y_2 &= \ka_{21}x_2 y_1 - d^*_2 y_2 -\ka_{12} x_1y_2  \label{xtalk4c}
\end{align}
Here, conditions for stability of the double extinction state are the same as above. Suppose $d_2\ka_{11}>d_1\ka_{21}$, so that if $\psi_{12} = 0$, this model behaves as the single metabolite model, and the state with $x_2 =0$, $x_1>0$ is stable. We are interested in causing the opposite extinction. That steady state is
\begin{equation}
    (x_1,x_2,y_1,y_2) = \left(0, \frac{1}{d_2}\left(f_1 - d_1^*\frac{d_2}{\ka_{21}}\right), \frac{d_2}{\ka_{21}}, \frac{1}{d_2^*}\left(f_1 - d_1^*\frac{d_2}{\ka_{21}}\right)\right)
\end{equation}
and the eigenvalues of the Jacobian matrix at this state can be computed symbolically, and the relavant eigenvalue is
\begin{equation}
    \lambda = \ka_{11}\frac{d_2}{\ka_{21}} - d_1 + \psi_{12}\left(f_1 - d_1^*\frac{d_2}{\ka_{21}}\right)
\end{equation}
giving a condition for stability on $\psi_{12}$ that can be achieved.

For coexistence, we will assume that the initial model with $\psi_{12} = 0$ has survival of $x_2$, so  $d_2\ka_{11}<d_1\ka_{21}$. Then we simply repeat the argument above to \emph{destabilize} the equilibrium point, causing $\lambda >0$. Then, all three of the double extinction and both single extinction equilibrium are unstable. We can conclude coexistence.

\end{appendices}

\end{document}